\documentstyle[aps,prl,twocolumn,epsfig]{revtex}

\draft
\begin{document}
\title{Spectral Compressibility at the Metal-Insulator Transition
\\of the Quantum Hall Effect
\thanks{Work performed within the
    research program of the Sonderforschungsbereich 341,
    K\"oln-Aachen-J\"ulich}
}
\author{Rochus Klesse and Marcus Metzler} \address{Institut f\"ur
  Theoretische Physik der Universit\"at zu K\"oln\\ D-50937 K\"oln,
  Germany\vspace{18pt}}
\author{\small\parbox{14.1cm}{\small
The spectral properties of a disordered electronic system at the
metal-insulator transition point are investigated numerically. 
A recently derived relation between the anomalous diffusion
exponent $\eta$ and the spectral compressibility
$\chi$ at the mobility edge, $\chi=\eta/2d$, is confirmed for 
the integer quantum Hall delocalization transition.
Our calculations are performed within the framework of an unitary
network-model and represent a new method to investigate spectral
properties of disordered systems.
}}
\address{\vspace{-15pt}}
\maketitle
Since the development of the classical Random Matrix Theory (RMT) by Wigner,
Dyson, Mehta and others the statistics of energy levels of complex
quantum systems have 
become the subject of research in many areas of physics, from the study of 
atomic nuclei to the investigation of disordered metals and quantum
chaos \cite{rmt}. The remarkable 
property the level statistics of these systems have in common is their 
universality, regardless of the microscopic details of the particular system.

In the following we will focus on the spectral properties of systems
undergoing a metal-insulator transition. In this case 
at least two entirely different types of level statistics are
involved.
In the insulating phase the probability density of an eigenstate is 
almost completely localized within a comparatively small volume.
As long as the spatial extension of any two states is small compared
to their distance, they are independent in the same way as two states
of two separate systems. 
Consequently, in the localized regime the energy levels are uncorrelated 
and therefore governed by Poisson statistics. For example,
the probability to find two consecutive levels separated by an energy
$\epsilon = s \Delta$,  $\Delta$ being the average level spacing,
is given by the Poisson-distribution $P(s) = \exp(-s)$.
The number variance 
$\Sigma_2(N) \equiv \langle (n - \langle n \rangle )^2 \rangle $ of an
energy interval which on average contains $N=\langle n \rangle$
levels is $\Sigma_2(N)=N$, according to the central limit theorem.

In the metallic or delocalized phase, on the other hand, the eigenstates are 
extended over the entire system. In this case the disorder potential causes
the levels to repel each other. This level repulsion leads
to a considerable rigidity of the spectrum with respect to fluctuations
in the level density: the number variance increases only 
logarithmically with the number of levels, $\Sigma_2(N) \propto
\ln(N)$ and the spectral compressibility defined by
\begin{equation} \label{chi}
  \chi=\lim_{N\rightarrow\infty}\lim_{L\rightarrow\infty}{d\Sigma_2(N)\over
     dN},
\end{equation}
vanishes; in the delocalized regime the spectra are incompressible.
The level spacing distribution is described by the so called 
Wigner surmise, $P(s) \propto s^\beta \exp(-c_\beta s^2)$, where
$\beta$ is a number of order one which depends on the symmetry of
the system's Hamiltonian \cite{rmt}. 
The factor $s^\beta$, missing in the Poisson statistics, reflects
the strong level-repulsion.

At the mobility edge, where the two phases with
their different kinds of spectra meet, things become more
complicated and have given rise to extended investigations and controversial
discussions concerning the critical level statistics 
\cite{shklovskii,altshuler,kravtsov_1994,aronov,kravtsov_1995,chalker_condmat,braun}. 
In the vicinity of the transition energy the probability density 
of a state is neither localized on a small confined area nor smeared
out almost homogeneously over the whole system, but forms a
self-similar measure which fluctuates very strongly on all
length scales. It is best described in terms of multifractality
\cite{multifraktales}. Since the repulsion between levels is
strongly influenced by the spatial correlations of the corresponding
eigenstates, their multifractal structure enters the level distribution.

In recent publications Chalker, Lerner and Smith presented a treatment
on level distributions
in disordered systems, which in contrast to earlier works
takes care of a possible non-trivial structure of the eigenstates
\cite{chalker_lerner}.
A central result of their work is a relation between the spectral
form factor $K(t)$ (i.e.\ the Fourier transformed of the
two-level-correlation function $R(s)$ )
and the ensemble averaged quantum return probability 
$p(t)$ of a state initially confined to a small volume.
Using this result and from
scaling theory that $p(t) \propto t^{-D_2/d} $, $D_2< d$ being the
fractal (correlation) dimension \cite{scaling},  Chalker et
al. \cite{chalker_condmat} derived the critical spectral compressibility 
\begin{equation} \label{result}
\chi = {\eta \over 2d} \:\: (<1),
\end{equation}
where the anomalous diffusion exponent $\eta$ is related to $D_2$ by
$\eta=d-D_2$ \cite{janssen_thesis}.

In spite of the Poisson-like behavior
of the number variance $\Sigma_2(N)$, there still is a strong 
level repulsion for small level separations. This is obvious 
in the spacing distribution function $P(s)$, which deviates 
only in the tail from the Wigner surmise 
\cite{shklovskii,kravtsov_1994,braun,critical_pvons} (Fig. \ref{fig-svonn}).
Hence, the non-vanishing compressibility must be due to level density
fluctuations on larger level separations.

Up to now, Eq.~(\ref{result}) could not be confirmed directly by numerical
calculations.
The numerical results achieved for the Anderson transition in $d=3$ 
dimensions \cite{altshuler,braun} are compatible with a
linear increase of the
number variance with $N$ but their rather large
numerical uncertainties do not allow to strictly exclude other 
dependencies. In particular, the slope of $\Sigma_2(N)$ for large $N$
could not be determined precisely. Besides,
the value of the multifractal exponent $D_2$ for the 3d-MIT is
not known very accurately \cite{3dd2}. In case of the MIT in
two-dimensional quantum Hall systems the only numerical simulation
\cite{feingold} presented so far shows no significant linear 
contribution to $\Sigma_2(N)$. 

The difficulties in determining the compressibility are
due to the necessity to investigate level separations that are large
compared to the average level spacing. 
Since only the critical region (which is usually a small
portion of the entire spectrum) can be used for the statistics,
one has to go to very large system sizes to produce a 
sufficient number of critical levels.
Unfortunately, the widths of the critical region achievable with present
computer capacities extends over no more than a few hundred levels,
so that the results are strongly affected by incalculable finite size 
effects. This can of course not be compensated by a large number of
disorder realizations.
Consequently, to study critical level statistics numerically
another way has to be found to produce larger critical spectra.

In this communication we show that by using the Chalker-Coddington
network model \cite{chalker_coddington}
for the Quantum Hall Effect the problems we just mentioned can be avoided.
As pointed out by Fertig in a detailed semiclassical analysis leading
to a description very similar to that in \cite{chalker_coddington}, an
energy dependent unitary operator $U(E)$ associated to the network
model offers an alternative method for the numerical calculations of
energy spectra and eigenstates \cite{fertig}. However, here we do not
use $U(E)$ for determining the real energy spectrum $E_n$, but
calculate the eigenvalues $e^{i\omega_l}$ of $U(E)$ itself, whereby
the energy $E$ is fixed to the critical value $E_c$. The obtained
quasi-spectrum 
$\omega_l$ is governed by the same statistics as the energy
spectrum at $E_c$ and is therefore suitable for our purposes.
By using this new method it is possible to improve the statistics
considerably and to confirm that $\chi= \eta/2 d$ to a high 
degree of accuracy.

Our numerical approach is closely related to that of Edrei et
al. \cite{edrei}, where the concept of a network model has
been used for calculating wave propagation through random
media. Actually, the definition of network states and operator used
here are in principle identical to those in \cite{edrei}. However, the
main difference are the boundary conditions. Whereas in
\cite{edrei} open systems are treated, since they focused on
transmission amplitudes, here we use closed systems in order to get
information about the energy level distribution.

The network-model for the Quantum Hall Effect
\cite{chalker_coddington} is based on ideas
developed in the early eighties \cite{anderson_shapiro} for the
description of the Anderson transition in terms of scattering theory.
It provides a semiclassical description of a
2d electron in a quantizing magnetic field and a smooth disorder
potential with correlation length $\lambda$ large compared to the
magnetic length $l_c$. The electron
executes a fast cyclotron 
motion on a circle of radius $l_c$ around a guiding center, which drifts
slowly along a contour $r$ of constant energy, $V(r)= E \equiv E' - 
\hbar \omega_c/2$. At saddle-points of the potential with energies close
to $E$ the electron tunnels with an appreciable probability between
different contours.

The motion of electrons along the contours is depicted by
one-dimensional, unidirectional channels, called links. The electron
tunneling between them is represented by $ 2\times 2$ scattering
matrices $S = \{t_{ml}\}$, which connect the complex current amplitudes
$\psi_k, \psi_l$ of
incoming links to those in outgoing links $\psi_m, \psi_n$ (Fig.~\ref{fig-u}). 
Due to the random length of the links between
neighboring saddle-points an electron acquires random
phases $\phi_j$, which we absorb into the
scattering coefficients $t_{ml}$.

The coefficients $t_{ml}$ depend on the electron energy $E$ and can, in 
principle, be determined by semiclassical methods for a given disorder
potential \cite{fertig}.
For a saddle-point at zero energy and tunneling energy $E_t$ the
tunneling amplitudes are $T = |t_{mk}|^2 = |t_{nl}|^2 = (1+
\exp(-E/ E_t))^{-1}$ and $R = |t_{ml}|^2 = |t_{nk}|^2 =
1-T $. Moreover, the random phases $\phi_{ml} \equiv
\arg(t_{ml})$ depend in a rather complicated manner on the energy $E$
\cite{fertig}.

A state $\Psi$ of the network is given by its complex amplitudes
$\psi_j$ on the links, $ \Psi=\{ \psi_j \}_j$. It is stationary at
energy $E$ if the scattering condition
\begin{equation}
  \psi_m = t_{mk}(E)\psi_k + t_{ml}(E)\psi_{l} \label{scattering_condition}
\end{equation}
is satisfied at each saddle point.

Defining an unitary operator $U(E)$ by
\begin{equation} \label{weq}
  U(E)e_l = t_{ml}(E)e_m + t_{nl}(E)e_n,
\end{equation}
where $e_i = \{ \delta_{ji} \}_j $, this condition can be written as
\begin{equation} \label{wpsi}
  U(E)\Psi = \Psi.
\end{equation}
The energy enters parametrically via the coefficients $t_{ml}(E)$. 
This equation has non-trivial solutions
only for discrete energies $E_n$. According to Fertig \cite{fertig},
these energies $E_n$ are eigenenergies of the system and their
eigenvectors $\Psi_n$ determine the amplitudes of the corresponding
eigenstates on the different equipotential contours (links).
Hence, under certain circumstances the scattering condition
(\ref{wpsi}) offers an alternative method of determining numerically
eigenenergies and states of an electron in an disordered system, as it
was first --- to the best of our knowledge --- pointed out by Fertig.
This method was utilized in \cite{klesse_metzler} for the numerical
calculation of critical eigenstates in quantum Hall systems.
(Like in \cite{edrei}, Eq. (\ref{wpsi}) can also be taken as a definition
of a time evolution for states at energy $E$, $\Psi(t+\tau) \equiv
U(E) \Psi(t)$, whereby the energy dispersion is neglected 
\cite{rk_thesis,bh_rk}.)

Before we proceed it is necessary to consider the length and energy 
scales determining the critical region. The averaged level spacing
$\Delta$ is related to the system size $L$ and magnetic length
$l_c$ via $\Delta \sim \Gamma (l_c/ L)^2$, $\Gamma$ being the width of
the disorder broadened Landau band. Close to the critical energy
$E_c=0$ the localization length $\xi$ is $\xi \sim \lambda
|E/\Gamma|^{-\nu}$. Therefore, the width of the critical energy region
is $\Delta_c \sim \Gamma (\lambda / L ) ^{1/\nu}$. Consequently, the
number $N_c$ of critical levels behaves like $N_c = \Delta_c /\Delta
\sim (\lambda/l_c)^2 (L/\lambda)^{2-1/\nu}$.

Note that even for a fixed ratio $L/\lambda$ the number $N_c$ can be 
enhanced by increasing the ratio $\lambda/l_c$.
We emphasize that in the Chalker-Coddington model this ratio is arbitrarily
large, so that the 
number of critical levels is not restricted, which makes the model 
in principle very convenient for investigations of critical level statistics.
However, the calculation of the energies $E_n$ by
solving the non-linear Eq. (\ref{wpsi}) is a difficult numerical task
and not suitable for practical purposes.

Therefore, in order to determine spectral statistics let us discuss 
instead of Eq. (\ref{wpsi}) the eigenvalue problem
\begin{equation} \label{eve}
U(E) \Psi_l(E) = e^{i \omega_l(E)} \Psi_l(E).
\end{equation}
For a given energy $E$ the unimodular eigenvalues
$e^{i\omega_l(E)}$ define quasi-energies
$\omega_l(E)$, $l=1,\dots,M=\dim U(E)$. They are smooth functions of the
energy and do not cross each other, following a theorem by von Neumann
and Wigner \cite{neumann}.
According to Eq.~(\ref{wpsi}), the intersection points of the curves
$\omega_l(E)$ with the lines $\omega = 0, \pm 2\pi, \pm 4\pi, \dots$
determine the energy levels $E_n$.

The flow of the levels $\omega_l(E)$ obeys two symmetries: First, the
intersection points $E_n'$ with shifted lines
$\omega'=\Omega,\Omega\pm 2\pi, \dots $ must exhibit the same
statistics as the original spectrum $E_n$, since the corresponding
transformed operator $U'(E)=e^{-i\Omega}U(E)$ belongs to the same
universality class as $U(E)$ ($U'$ deviates from $U$ only by a global
phase shift, which has no influence on the statistical
properties). Second, as long as the critical regime is not left,
$|E|<\Delta_c$, the statistical properties of the flow $\omega_l(E)$
can not change significantly with energy, because such a change would be
accompanied by a new energy scale inbetween $\Delta$ and $\Delta_c$
($ \ll E_t \ll \Gamma$), which makes no physical sense.

Due to this homogeneity in both directions and due to the strong
repulsion of the $\omega_l(E)$ they must behave as depicted in
Fig.~\ref{fig-flow}: The average slope of the curves varies neither
strongly with 
the level number $l$ nor with the energy $E$. Further, this homogeneity
implies that the intersection points $\omega_l^c$ with a cut $c$
crossing the band of curves show essentially the same statistics,
independent of the precise position of $c$. For this reason,
instead of the real energy spectrum $E_n$  one can also use 
a quasi-spectrum $\omega_l(E)$ with $E$ within the critical regime for
an analysis of the critical level statistics \cite{jalabert}.
A big advantage of this method is the 
simple fact that the $\omega_l(E)$ are far better numerically
accessible than the real energies $E_n$. They can be calculated by
solving the linear eigenvalue problem (\ref{wpsi}) with standard
numerical methods. \cite{erlaeuterung}

For our calculations we used closed networks of $50 \times 50 $ saddle-points
with periodic boundaries in one and reflecting in the other
direction. The transmission amplitudes were set to the critical value
$ T = 1/2$ for models describing the transition point and to 
$ T_\pm= (1 + \exp(\pm E/E_t))^{-1}$ with $E/E_t=10$ for non-critical
systems with strongly localized states. The disorder is represented
by random scattering phases $\phi_{ml} = \arg(t_{ml})$. 
Note that the
calculations are done at constant energies $E=E_c$ and $E=10E_t$,
respectively, hence we did not have to know the
energy dependence of the phases $\phi_{lm}(E)$.
From this
settings we obtained random network operators at the critical point,
$U(E=E_c)$, and deep in the localized regime, $U(E=10E_t)$, of
dimension $M=2\times 50\times 50$. Diagonalizing them by standard
numerical methods yields critical ($E=E_c$) and non-critical
($E=10E_t$) quasi-spectra of $M=5000$ levels each. 

As explained in the considerations given above, the critical level
statistics can be determined by analyzing the critical
quasi-spectra. Although the statistics do not change within the
quasi-spectra one has to take into account that the total number of
quasi-levels per spectrum is fixed to $M$. So, when calculating the
number variance $\Sigma_2(N)$ one has to confine oneself to 
interval sizes $\Delta\omega$ with averaged level number $N=\langle n
\rangle_{\Delta\omega}$ small compared to $M$. We checked by numerical
simulations with Poisson distributed levels that at a total number of
$M=5000$ levels deviations from the expected number variance
$\Sigma_2(N)=N$ are negligible for $N < 300$.

For the determination of the critical number variance $\Sigma_2(N)$ we
divided the quasi-spectra of 40 different disordered critical network
operators $U(E_c)$ into non-overlapping intervals of length
$\Delta\omega=(2\pi/5000)N$, $N$ ranging from 1 to 300. For each
$N$ this results in an ensemble of $m_N = 40 \times M /N$ intervals
with level numbers $n_i$ and $\langle n_i \rangle = N$, from which we
calculate the level 
number variance $\Sigma_2(N) = m_N^{-1}\sum_{i=1}^{m_N} (n_i-N)^2$.

The results plotted in Fig.~\ref{fig-svonn}
show a clearly linear behavior of the number variance $\Sigma(N)$ for
a wide range of $N$.
Eq.~(\ref{result}) predicts a slope
of $\eta /2d = (2-D_2)/2d = 0.125 \pm 0.01$, where we have taken
$D_2=1.5 \pm 0.05$  from independent numerical calculations of
critical states \cite{dezwei,mocm_thesis,rk_thesis}.
A least square fit of our data yields
a slope $0.124 \pm 0.006$, which agrees with the prediction very well.

The dashed lines mark the range of the expected fluctuations of
$\Sigma_2(n)$ due to the finite number $m_N$ of intervals, calculated
via the $\chi^2_\alpha$-distribution for $\alpha = 0.8, 0.2$. This
clearly indicates that the deviations of the data from the straight
line are not systematic but due to statistical fluctuations.

The spacing distribution $P(s)$ plotted in the inset of Fig.~\ref{fig-svonn}
shows for small spacings $s$ a WD-type behavior for a GUE ensemble
(dashed line), $P(s) \propto s^2$, indicating strong level repulsion for small 
level spacing.

We use the same procedure as for the critical quasi-spectra
for five non-critical quasi-spectra at $E= 10E_t$ in the strongly
localized regime. Here $\Sigma_2(N)$ follows a straight line of slope
1, as it should be, since in this region the levels are
Poisson-distributed. 

To summarize, the recently derived relation between the multifractal
exponent $\eta= d-D_2$ of eigenstates and the spectral compressibility
at the mobility edge, $\chi=\eta/2d$, has been confirmed numerically for the
integer quantum Hall delocalization transitions. This has been done
by introducing a new method to investigate spectral properties
of disordered systems.

We would like to thank J\'anos Hajdu, Bodo Huckestein and Martin Janssen
for valuable discussions and the research program Sonderforschungsbereich
341, K\"oln-Aachen-J\"ulich for their support.

\begin{figure} 
  \begin{center}
  \end{center}
  \caption{The Chalker-Coddington network. At each saddle point a
    scattering matrix $S$ describes the transition from incoming to
    outgoing states. The operator $U$ maps each incoming link
    amplitude to the two outgoing links according to the transmission
    coefficients $t_{ml}, t_{nl}, \dots$.\label{fig-u} }
\end{figure}

\begin{figure}
  \begin{center}
  \end{center}
  \caption{ The eigenvalues $\exp(i\omega_l(E))$ of $U(E)$ as functions
      of the energy $E$. The intersections with the lines $\omega=
      2\pi z, z \in Z$ determine the energy-levels $E_n$,
      those with the $E=E_c $-line the quasi-energies
      $\omega_l(E_c)$.\label{fig-flow}} 
\end{figure}

\begin{figure}    
  \begin{center}
  \end{center}
  \caption{The level number variance  $\Sigma_2$ plotted against the average
    number of levels $N$ at the critical point ($\Diamond$) and in
    the localized regime ($\Box$). The solid line has a slope of $0.124$,
    the dashed lines mark the range of the expected statistical fluctuations
    ($\chi^2_\alpha$, $\alpha=0.8,0.2$). 
    The dotted line has a slope of $1$. The inset shows the level 
    spacing distribution $P(s)$ at the critical 
    point ($\Diamond$) and the WD-distribution for GUE (dashed line).
    \label{fig-svonn}}
\end{figure}


\begin{thebibliography}{10}
\bibitem{erlaeuterung}
In order to get a better understanding of the quasi-energies
$\omega_l$ and their 
eigenfunctions $\Psi_l$ we refer to recent numerical works
\cite{rk_thesis,mocm_thesis,bh_rk}.
In dynamical simulations \cite{rk_thesis}, where $U(E_c)$ is interpreted
as evolution operator 
for microscopic time steps --- this means that the $\omega_l(E_c)$
are thought of as actual eigenenergies $E_l$ with eigenstates $\Psi_l(E_c)$
---, the well known critical anomalous diffusion behavior shows up in
good agreement with the predictions of scaling theory \cite{scaling}.
Moreover, the correlations in the local amplitudes of eigenfunctions
$\Psi_l(E_c)$, $\Psi_{l'}(E_c)$ were found to obey the same scaling
behavior in the spatial and quasi-energy difference $\Delta \omega =
|\omega_l-\omega_{l'}|$ as those of real critical eigenfunctions in
real energy differences \cite{mocm_thesis,bh_rk}.
This suggests to interpret the quasi-spectrum $\omega_l(E)$ of $U(E)$
as a spectrum, which is statistically equivalent to the
excitation spectrum of the real system in the vicinity of the energy $E$.
\end{thebibliography}
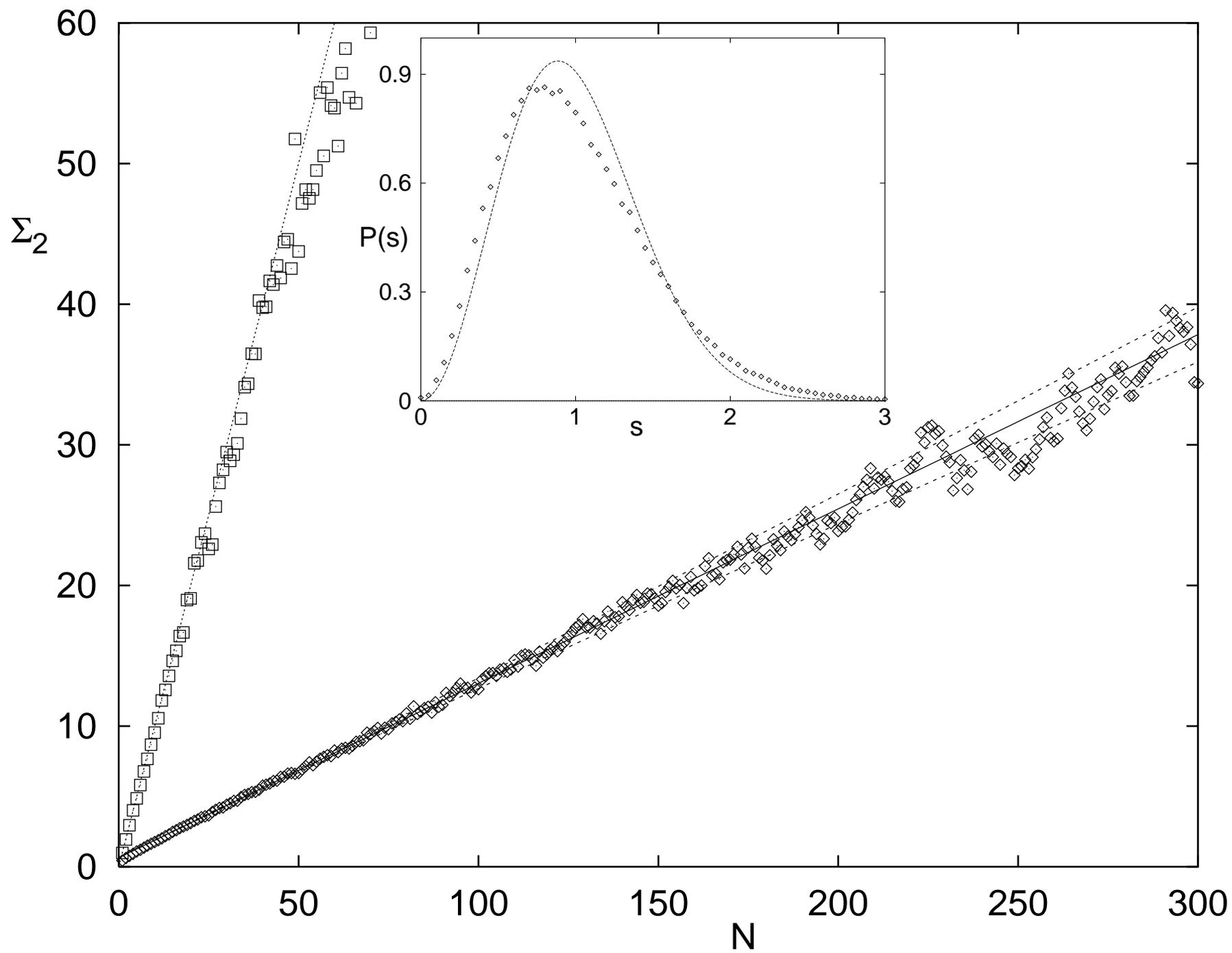
\end{document}